\documentclass[a4paper,11pt]{article}
\pdfoutput=1 

\usepackage{jinstpub} 
\usepackage{subcaption}

\title{\boldmath Radon mitigation during the installation of the CUORE $0\nu\beta\beta$ decay detector}

\author[a,b]{G. Benato,}
\author[b,c]{D. Biare,}
\author[c]{C. Bucci,}
\author[b,c]{L. Di Paolo,}
\author[a,b\phantom{,}1]{A. Drobizhev,\note{Corresponding author.}}
\author[a,b]{Yu.G. Kolomensky,}
\author[d\phantom{,}*]{R.W. Kadel,\note[*]{Retired}}
\author[e]{J. Schreiner,}
\author[a,b]{V. Singh,}
\author[f]{T. Sipla,}
\author[f]{J. Wallig,}
\author[f]{and S. Zimmermann}

\affiliation[a]{\textit{Department of Physics, University of California, Berkeley, CA 94720, USA}}
\affiliation[b]{\textit{Nuclear Science Division, Lawrence Berkeley National Laboratory, Berkeley, CA 94720, USA}}
\affiliation[c]{\textit{INFN Laboratori Nazionali del Gran Sasso, Assergi (AQ), 67100, Italy}}
\affiliation[d]{\textit{Physics Division, Lawrence Berkeley National Laboratory, Berkeley, CA 94720, USA}}
\affiliation[e]{\textit{Max-Planck-Institut f{\"u}r Kernphysik, Heidelberg, 69117, Germany}}
\affiliation[f]{\textit{Engineering Division, Lawrence Berkeley National Laboratory, Berkeley, CA 94720, USA}}

\emailAdd{adrobizhev@lbl.gov}

\abstract{CUORE---the Cryogenic Underground Observatory for Rare Events---is an experiment searching for 
  the neutrinoless double-beta ($0\nu\beta\beta$) decay of $^{130}$Te with an array of 
  988 TeO$_2$ crystals operated as bolometers at $\sim$10 mK in a large dilution refrigerator. 
  With this detector, we aim for a $^{130}$Te $0\nu\beta\beta$ decay half-life sensitivity of 
  $9\times10^{25}$ y with 5 y of live time, and a background index of $\lesssim 10^{-2}$  
  counts/keV/kg/y. Making an effort to maintain radiopurity by minimizing the bolometers' exposure 
  to radon gas during their installation in the cryostat, we perform all operations inside a 
  dedicated cleanroom environment with a controlled radon-reduced atmosphere. In this paper, 
  we discuss the design and performance of the CUORE Radon Abatement System and cleanroom, 
  as well as a system to monitor the radon level in real time.}

\keywords{Gas systems and purification; Radiation monitoring}

\arxivnumber{1711.07936}

\begin{document}
\maketitle
\flushbottom

\section{Introduction}
\label{sec:intro}

The CUORE experiment at Laboratori Nazionali del Gran Sasso (LNGS) in Italy~\cite{Qahep,Qproposal} 
comprises 988 5$\times$5$\times$5 cm$^3$ TeO$_2$ crystals instrumented with neutron-transmutation-doped 
germanium thermistors (NTDs) and mounted in copper 
frames with PTFE holders. The crystals are arranged in 19 towers of 52 bolometers each. To maintain 
radiopurity, all assembly work is carried out in nitrogen-atmosphere overpressure glove boxes, 
making use of robots. Completed components are stored in nitrogen-flushed acrylic containers. 
All of the gloveboxes and storage are located inside the same class-1000 cleanroom facility.

\begin{figure}[htbp]
  \centering
  \begin{subfigure}[t]{0.5\textwidth}
    \centering
    \includegraphics[width=0.95\linewidth]{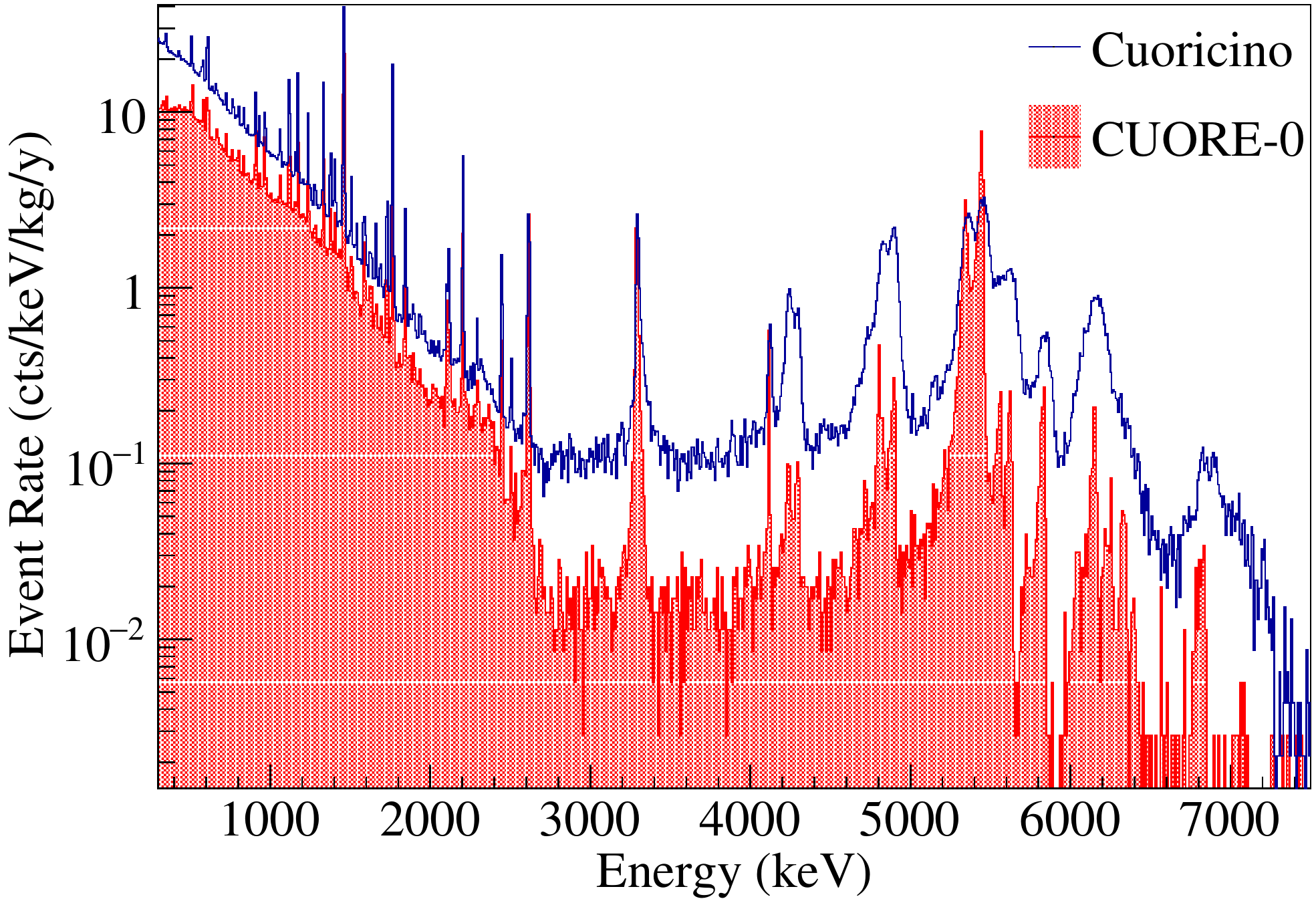}
    \caption{\label{fig:Q0_Qino_bgnds} } 
  \end{subfigure}%
  \begin{subfigure}[t]{0.5\textwidth}
    \centering
    \includegraphics[width=0.95\linewidth]{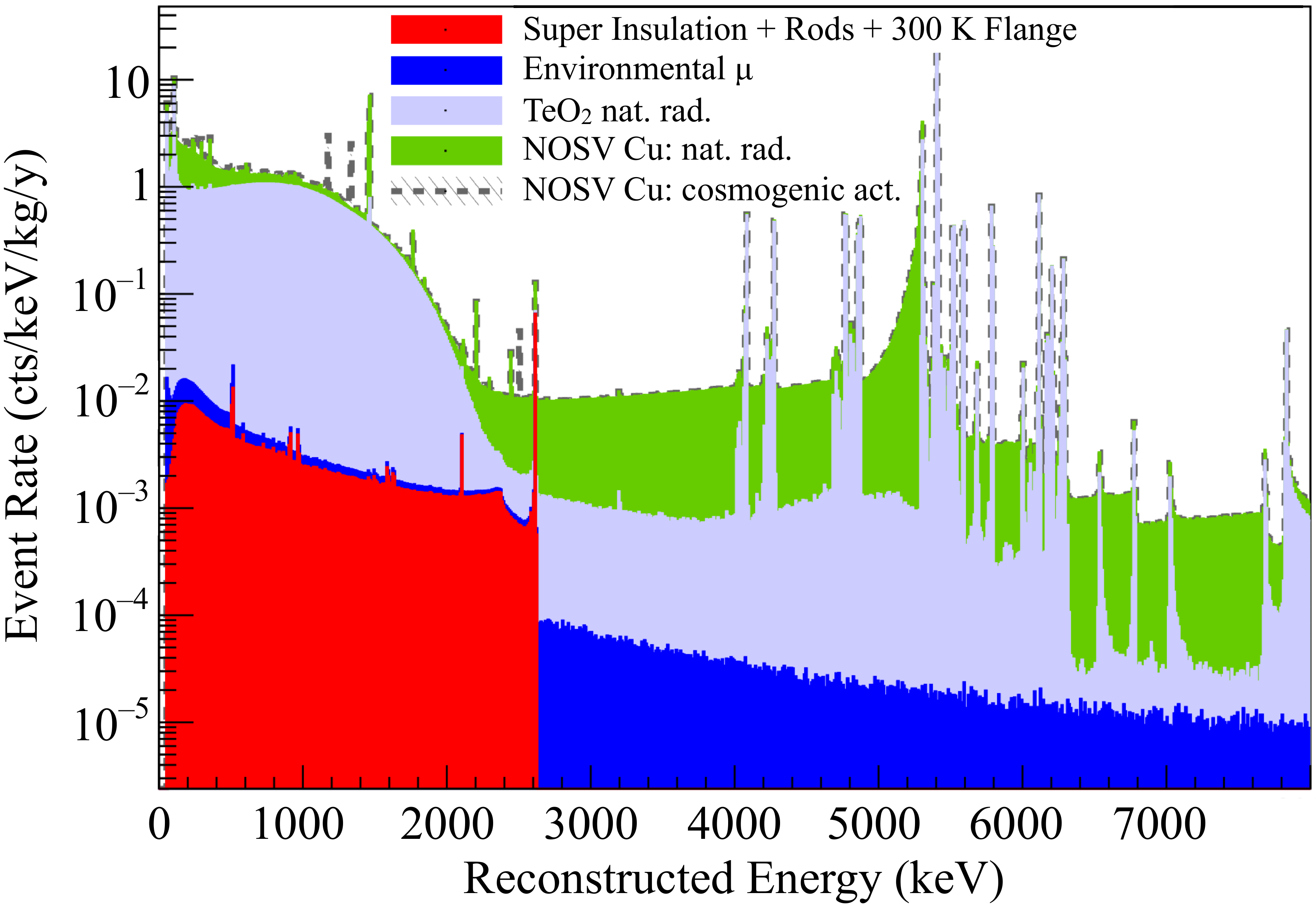}
    \caption{\label{fig:BgndBudget} }
  \end{subfigure}
  \caption{\label{fig:Qspectra} \footnotesize{\textbf{(a)} Comparison of CUORE-0 and Cuoricino spectra, demonstrating background 
      reduction. \textbf{(b)} Simulated CUORE energy spectrum from background budget study~\cite{QBB}. The peak at 
      $\sim$3.3 MeV in the experimental spectra comes from a $^{190}$Pt crystal bulk contamination, which 
      does not contribute to the background at Q$_{\beta\beta}$\cite{Q02nu}, and is not included in the 
      CUORE background budget.}}
\end{figure}

We first used the CUORE assembly line for the construction of the CUORE-0 demonstrator experiment,
which consists of a single CUORE-type tower~\cite{Q0performance,Q0detector}. CUORE-0 achieved a 
factor of $\sim$10 reduction in $\alpha$ background~\cite{Q0PRL,Q0PRC} 
over the Cuoricino prototype, though they were operated in the same cryostat  
(figure~\ref{fig:Q0_Qino_bgnds})~\cite{QinoNDBD,QinoResults}.

Though the CUORE detector's design and fabrication are identical to a scaled-up CUORE-0, its target
background in the region of interest around the $^{130}$Te Q-value (2.53 MeV) is lower by a factor 
of $\sim$6 (figure~\ref{fig:BgndBudget})~\cite{QBB,Q02nu}. This is possible in a new custom-built 
cryostat, made with clean copper and carefully selected materials~\cite{Cu}, as well as improved 
shielding. We need to ensure, however, that no significant recontamination occurs during the 
installation of the instrument in the cryostat. During this operation, the towers are removed from 
their nitrogen-flushed storage, and are exposed to the surrounding atmosphere until the cryostat's 
inner vacuum chamber (IVC) is evacuated.

Given that the towers are very fragile, $\sim$1 m tall, and weigh a total of nearly 1 t, we opt 
to perform this operation in an air, rather than nitrogen, atmosphere. A temporary 
higher-specification cleanroom enclosure (CR6) of $\sim$32 m$^3$ volume with a radon-free air supply 
is erected around the CUORE cryostat, within the confines of our standard class-1000 cleanroom (CR5).

Our goal, to avoid recontamination, is to keep the radon activity of the air below a 1 Bq/m$^3$ 
limit for the duration of the installation~\cite{Clemenza}. The clean air is supplied by the Radon Abatement System, 
which scrubs ambient air through carbon filters. Inside CR6, the air is circulated and cleaned of airborne 
dust by two ULPA (Ultra-Low Particulate Air) 
filter cabinets. The radon level is constantly monitored by an electrostatic radon monitor (RM).

\section{Description of the setup}
\label{sec:setup}

\subsection{Radon abatement system}
\label{subsec:ras}

\begin{figure}[htbp]
  \centering
  \begin{subfigure}[t]{0.6\textwidth}
    \includegraphics[width=1.0\linewidth]{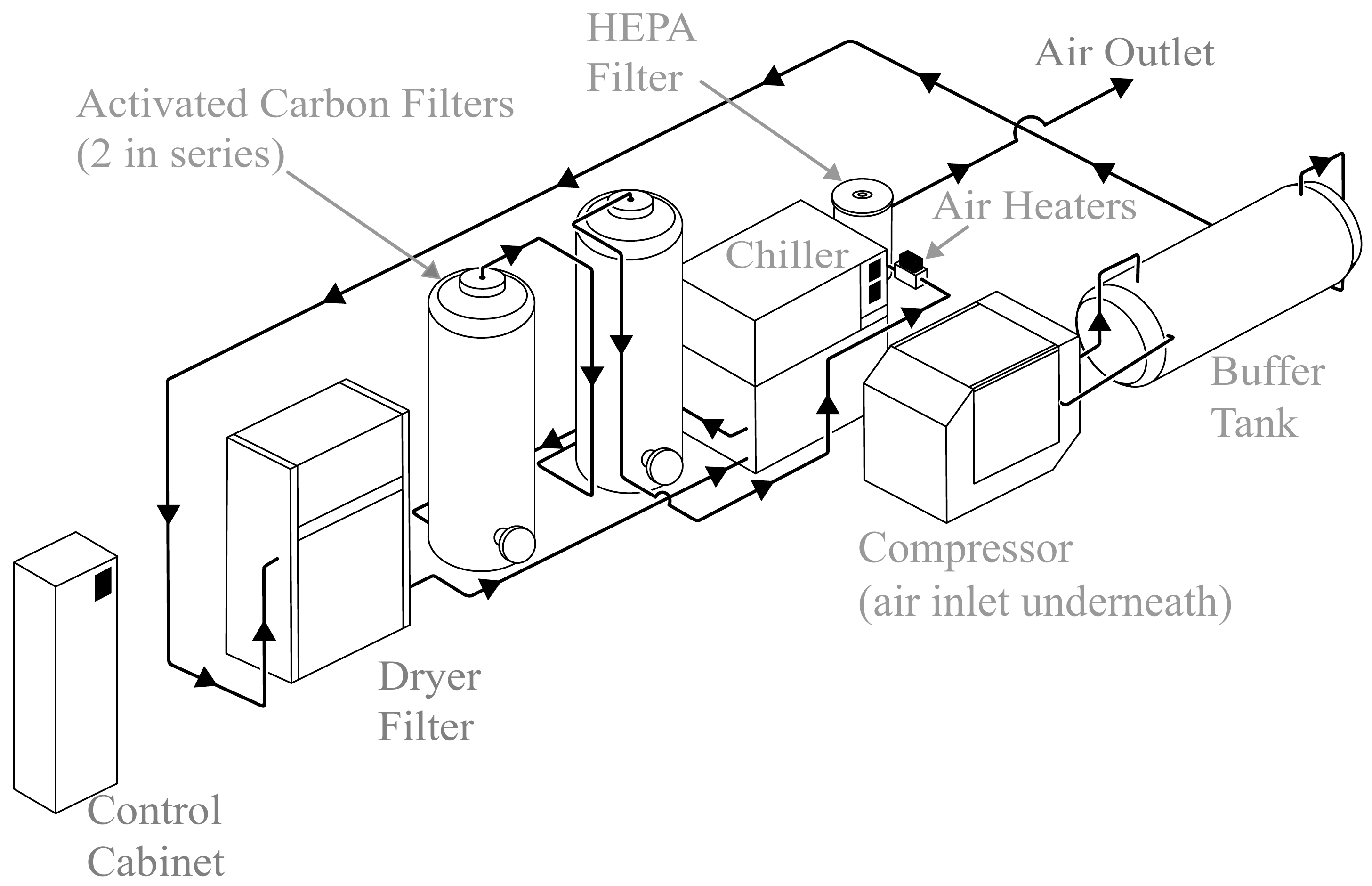}
    \caption{\label{fig:RASdiag} }
  \end{subfigure}%
  \begin{subfigure}[t]{0.4\textwidth}
    \includegraphics[width=0.95\linewidth]{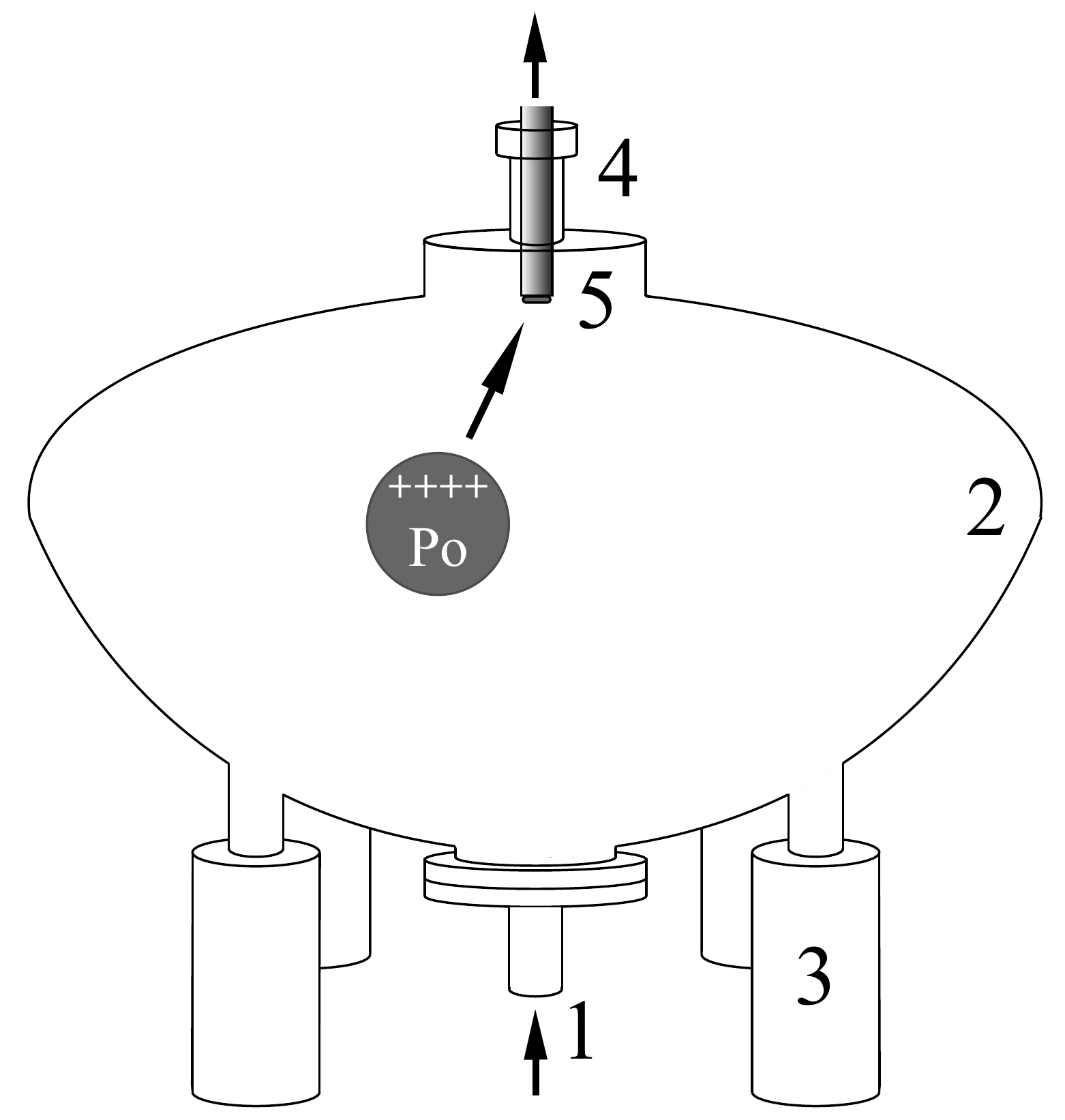}
    \caption{\label{fig:RM} }
  \end{subfigure}
\caption{\label{fig:diagrams} \footnotesize{\textbf{(a)} Diagram of the RAS, labeling key components, with ducting paths altered 
      and simplified for clarity. The physical arrangement shown corresponds to the CUORE setup. 
      The system is located adjacent to the cleanroom facility. \textbf{(b)} Schematic of the MPIK RM, showing the 
      gas inlet (1) of the $\sim$700 L vessel (2) on the bottom, and the outlet at the top (4). The vessel 
      sits at $\sim$10 kV on insulated feet (3). The Si PIN diode (5) is grounded and mounted on an insulated 
      cylindrical finger near the vessel outlet.}}
\end{figure}

The Radon Abatement System (RAS), manufactured by the Ateko company of the Czech Republic, functions by
cleaning air from the surrounding environment. Upon intake, air is pressurized by a compressor to $\sim$9 
atm, passed through an oil vapor separator and three microfilters, and fed into a dryer. 
There, the air is desiccated such that the dew point is below $-$70 $^\circ$C, and filtered again of any
eventual liquid or dust. This, in turn, permits us to cool the air down to $\sim -$55 $^\circ$C, prior to 
flushing it through two large activated carbon filters (in series) to trap the radon. The cooling is necessary, 
because activated carbon's efficiency for radon absorption decreases exponentially with temperature.
These filters are a potential source of dust, particularly problematic for cleanroom use, so we pass 
the air through a set of coarse particle filters followed by HEPA (High Efficiency Particulate Air) filters 
after heating it back to room temperature (between 14 and 25 $^\circ$C). Figure~\ref{fig:RASdiag} shows a 
diagram of the RAS. It is deployed on an external platform/balcony of the underground hut housing the CUORE 
experiment, adjacent to and one story above the cleanroom facility, allowing us to conveniently feed clean 
air through a port near the ceiling of the cleanroom.

At the output of the RAS, the air has a $^{222}$Rn activity reduced to $<$ 5 mBq/m$^3$ (from $\sim$30 
Bq/m$^3$ ambient), and is produced at a rate of $\sim$120 m$^3$/h.

\subsection{Radon monitor}
\label{subsec:rm}

\begin{figure}[htbp]
  \centering

\end{figure}

CUORE is vulnerable to contamination from the surrounding atmosphere over the course of the detector 
installation process, as well as the subsequent closing of thermal and radiation shields and other 
hardware work. This corresponds to a period of about four months, during which it is crucial for us 
to track the level of radon in the cleanroom, in case of failure or degradation of containment and/or 
the RAS. Given the generally low level of radon, the detector must be very sensitive. Additionally, 
because sudden failures are possible, and we need to be able to respond to those quickly, 
it must be fast. This fact prevents us from using conventional radon meters, such as the Durridge 
RAD7, which can reach a sensitivity below 1 Bq/m$^3$ only by integrating over a long period of time. 

Thus, we make use of a highly sensitive electrostatic radon monitor (RM) belonging to the Max Planck 
Institute for Nuclear Physics (MPIK) of Heidelberg, Germany. This is located at the facility of the 
GERDA experiment, also in Hall A of LNGS, about 80 m away from CUORE. The RM is comprised of a 700 L
vessel, through which the monitored gas is flushed. A silicon PIN diode is positioned at the outlet to 
maximize efficiency, at a 10 kV potential offset from the vessel (figure~\ref{fig:RM}). The RM has a static sensitivity of 
0.5 mBq/m$^3$, which improves to 50 $\mu$Bq/m$^3$ with an air flow rate of 7 L/min.

\begin{figure}[htbp]
  \centering
  \begin{subfigure}[t]{0.45\textwidth}
    \centering
    \includegraphics[width=0.95\linewidth]{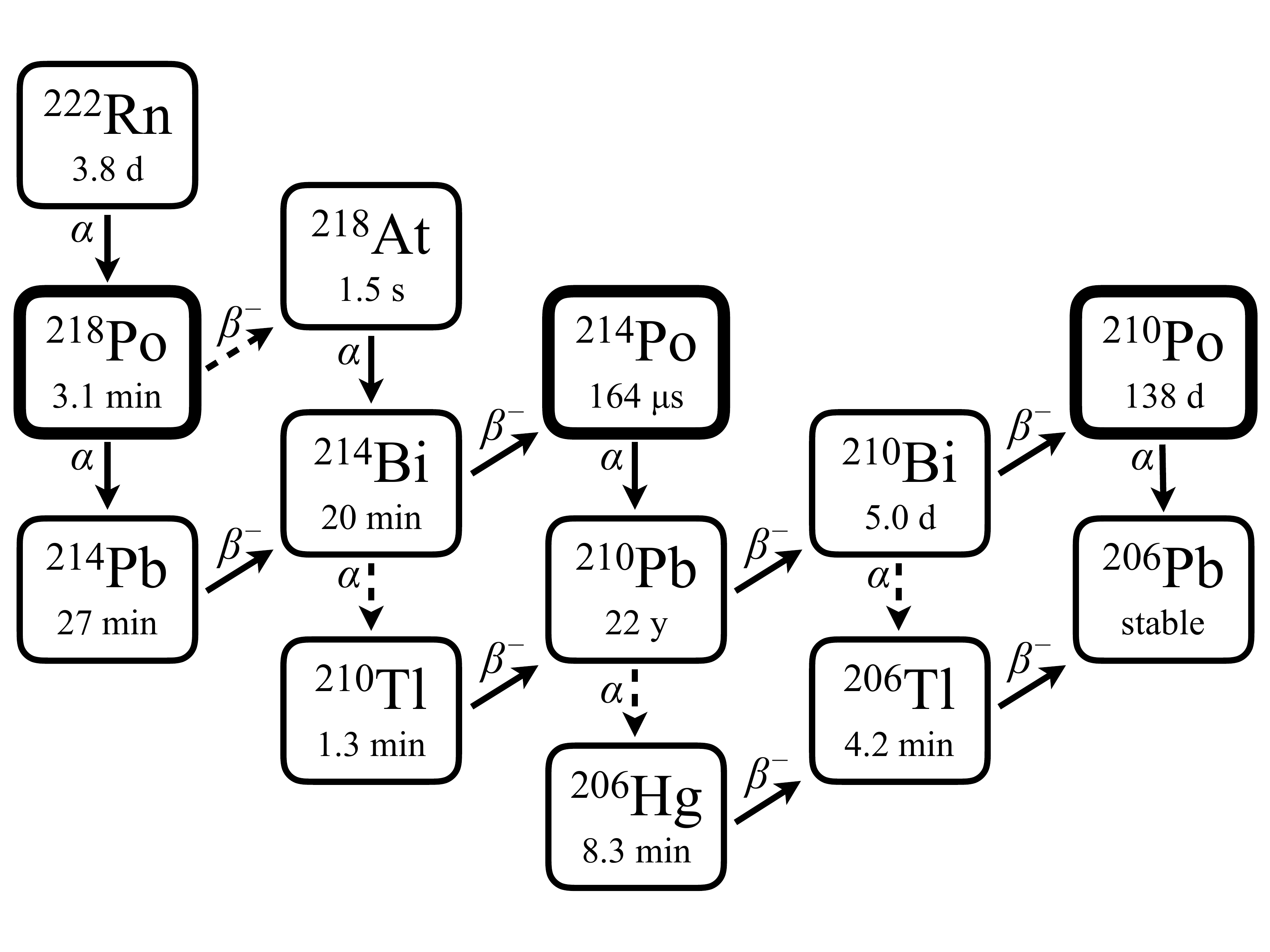}
    \caption{\label{fig:Uchain} }
  \end{subfigure}%
  \begin{subfigure}[t]{0.55\textwidth}
    \centering
    \includegraphics[width=1.0\linewidth]{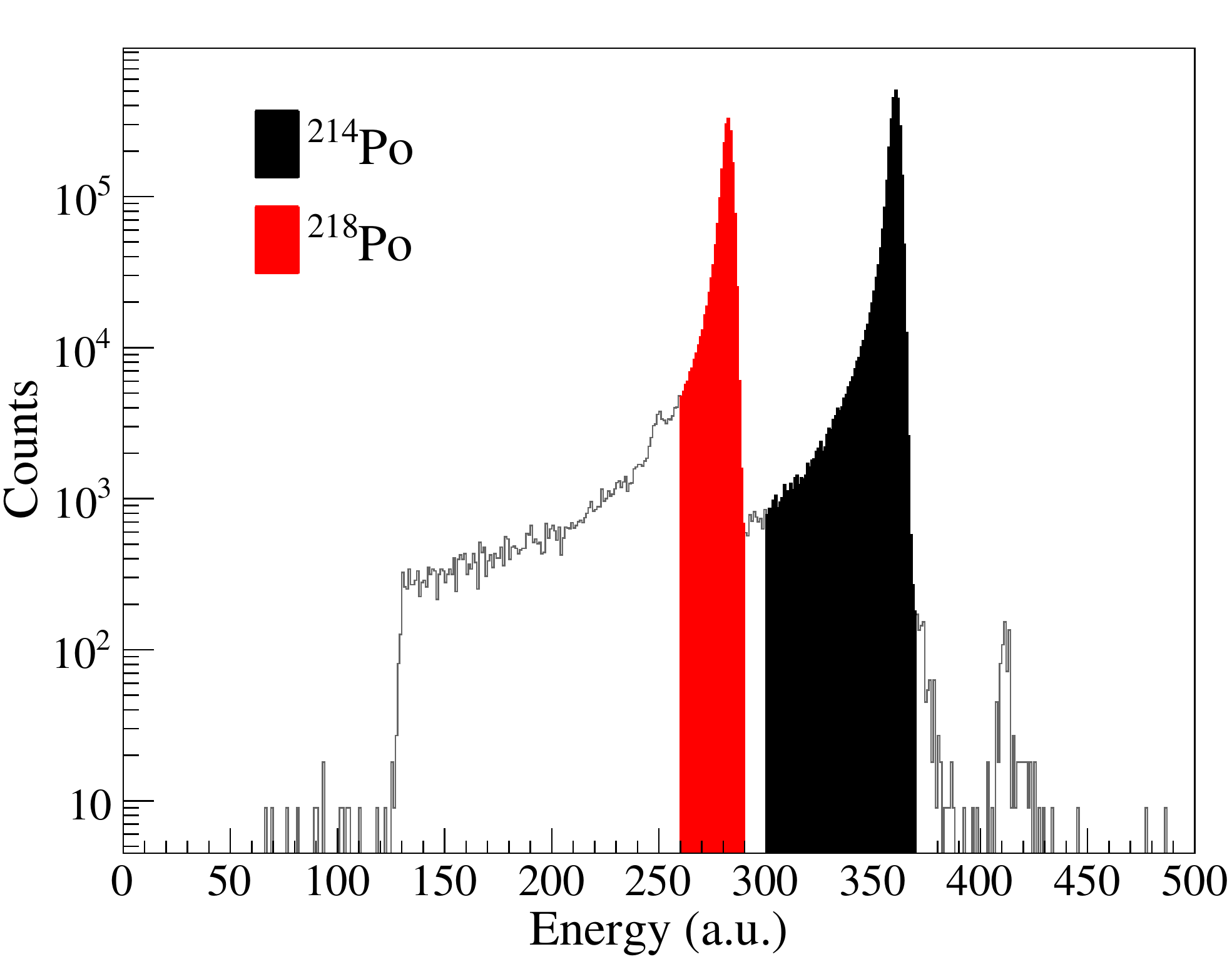}
    \caption{\label{fig:RMspectrum} }  
  \end{subfigure}
  \caption{\label{fig:Radon} \footnotesize{\textbf{(a)} The decay of $^{222}$Rn, from the $^{238}$U decay chain, 
    which is the primary source of the radon contamination we are mitigating. Solid arrows indicate
    dominant decay channels, while dashed arrows are secondary processes. The polonium
    isotopes we detect directly are highlighted in bold. \textbf{(b)} Radon Monitor energy spectrum: $^{218}$Po 
    (red) and $^{214}$Po (black) $\alpha$ peaks at 6.1 MeV and 7.8 MeV, respectively. Smaller peaks 
    from $^{210}$Po at 5.4 MeV and $^{212}$Po at 9 MeV are also visible.}}
\end{figure}

When a $^{222}$Rn nucleus $\alpha$-decays inside the RM, the $^{218}$Po daughter is positively ionized, 
and drifts along the voltage gradient towards the grounded diode. If this nucleus does not decay or 
recombine to become electrically neutral on the way, it adheres to the diode surface and decays there to 
$^{214}$Pb ($T_{1/2}^{218\mathrm{Po}} \simeq 3$ min.). The resulting $\alpha$ particle is detected with a 
$\sim$50\% geometric efficiency. Moving along the uranium chain (figure~\ref{fig:Uchain}), the $^{214}$Pb 
and its $^{214}$Bi daughter undergo relatively fast $\beta^{-}$ decay ($T_{1/2}^{214\mathrm{Pb}} \simeq 27$ min., 
$T_{1/2}^{214\mathrm{Bi}} \simeq 20$ min.) to produce another polonium isotope---$^{214}$Po. This $\alpha$ 
decay, to $^{210}$Pb ($T_{1/2}^{214\mathrm{Po}} \simeq 164$ $\mu$s), can also be detected by the diode. 
In fact, the probablility of observing the $^{214}$Po $\alpha$ is greater than that from 
$^{218}$Po: in the event that $^{218}$Po decays in the gas, all of the daughters mentioned above 
tend to keep moving in the direction of the diode, increasing the chance that the $^{214}$Po is 
resting on the diode surface at the moment of decay.

Thus, the experimental spectrum, shown in Figure~\ref{fig:RMspectrum}, is characterized by the 
$^{218}$Po $\alpha$ peak at 6.1 MeV and a somewhat taller $^{214}$Po $\alpha$ peak at at 7.8 MeV.
The value of the ratio between the observed peak amplitudes depends on the probability of 
recombination, and ultimately on the properties of the monitored gas. These data are consistent with
a total efficiency of $\sim$30$-$40$\%$ for $^{218}$Po and $^{214}$Po. As we can see 
in Figure~\ref{fig:Uchain}, the $^{238}$U chain also contains $^{210}$Po. Producing a 5.4 MeV 
$\alpha$, this isotope is the source of the small bump in the tail of the $^{218}$Po peak. 
Its amplitude is much smaller due to the longer half-lives involved, particularly for $^{210}$Pb
(about 22 y). The peak includes residual $^{210}$Pb from all past measurements made
by the meter. The contribution from the $^{232}$Th chain is much smaller. The 6.9 MeV $\alpha$ 
from $^{216}$Po is not visible in the spectrum beneath the $^{214}$Po tail, while $^{212}$Po 9 MeV
$\alpha$ is responsible for the residual peak near 415 a.u. in Figure~\ref{fig:RMspectrum}.

\subsection{Cleanroom}
\label{subsec:cr6}

\begin{figure}[htbp]
  \centering
  \begin{subfigure}[t]{0.5\textwidth}
    \centering
    \includegraphics[width=0.95\linewidth]{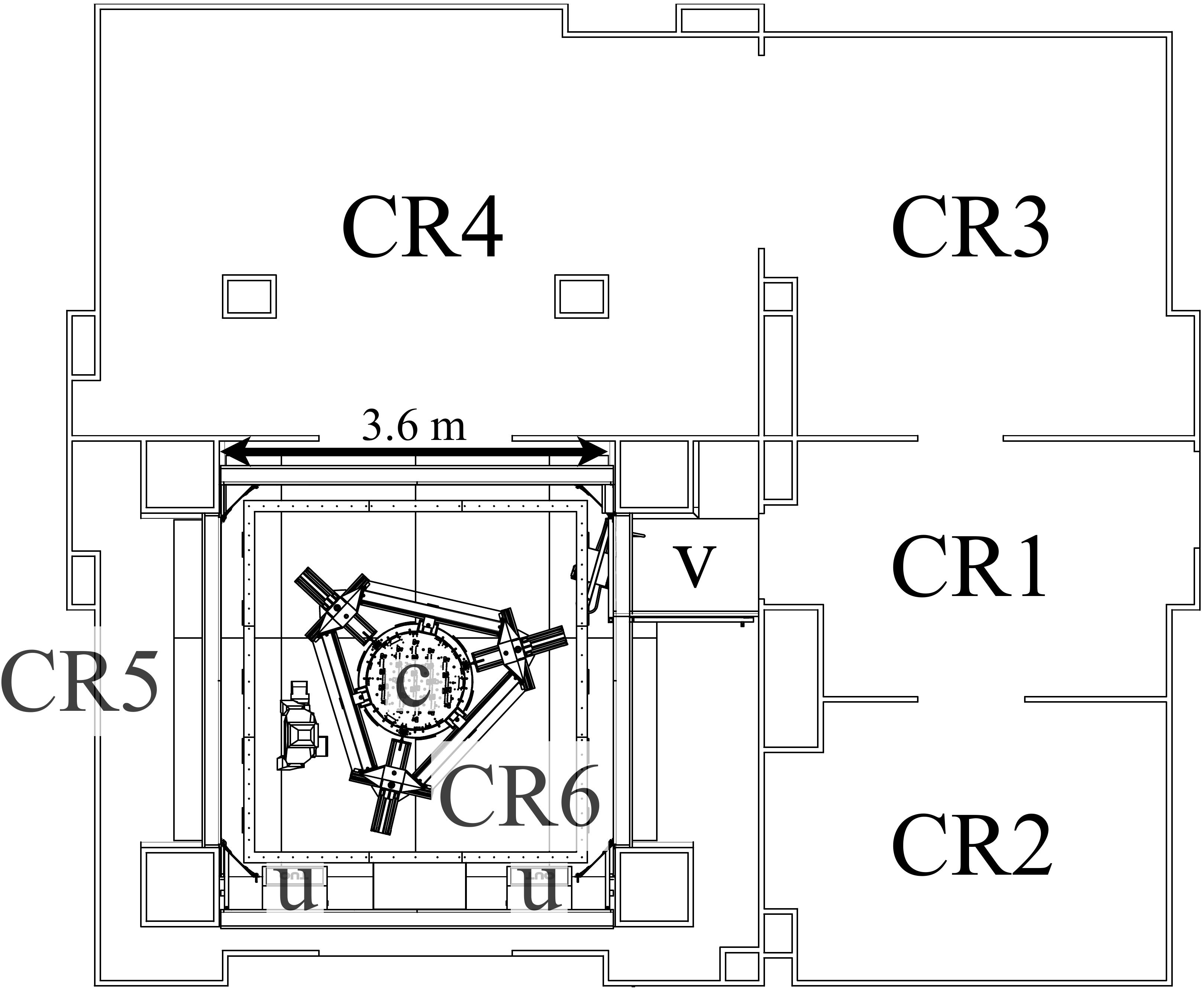}
    \caption{\label{fig:CRfull} }
  \end{subfigure}%
  \begin{subfigure}[t]{0.5\textwidth}
    \centering
    \includegraphics[width=0.95\linewidth]{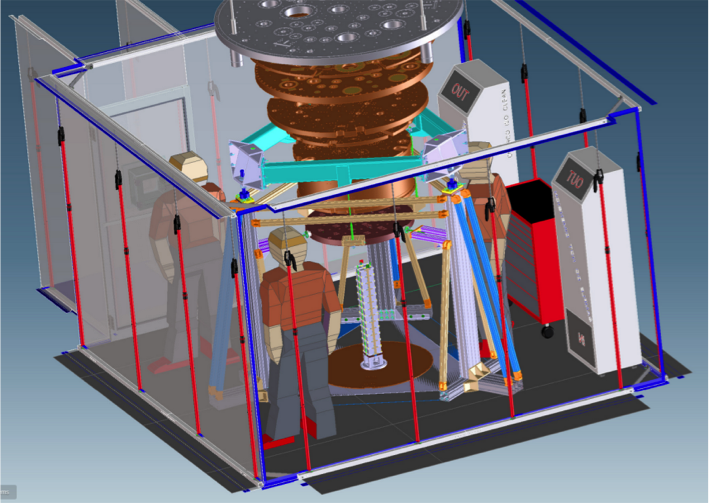}
    \caption{\label{fig:CR6} }
  \end{subfigure}
  \caption{\label{fig:CR} \footnotesize{\textbf{(a)} A line drawing of the CUORE cleanroom facility showing the 
    separate spaces: entry area (CR1), detector assembly (CR2, CR3), tower storage (CR4), and 
    the cryostat (c) room (CR5). The temporary cleanroom (CR6) for detector installation 
    is found within CR5, connecting to CR1 via the vestibule (v). Air from the RAS feeds into 
    CR6 from a port in the ceiling adjacent to CR4, and is recirculated through two ULPA filter
    cabinets (u). \textbf{(b)} Rendering of CR6, showing the soft walls, vestibule, 
    seals in blue, two ULPA filter cabinets, as well as the cryostat and detector installation 
    hardware inside.}}
\end{figure}

Our cleanroom facility is located on the second (middle) floor of the CUORE experiment's underground
hut. As originally constructed, it comprises five separate rooms (figure~\ref{fig:CRfull}: 
CR1 for entry, CR2 and CR3 for detector assembly, CR4 for storage of completed towers, and CR5 housing 
the cryostat). 

The assembly and storage areas are certified as class 1000, with a handler supplying ambient air 
(containing $\sim$30 Bq/m$^3$ of radon) through HEPA filters. This is sufficiently clean for our 
purposes, because the detector assembly gloveboxes and tower storage containers are all nitrogen-flushed.
CR5, housing the cryostat, is equivalent to the others, but additionally features a large 
trap door in the center for raising and lowering the thermal and radiation shields, and a 
double door to the hall for forklift access. This floor also saw a significant amount of heavy 
equipment operations over the course of $\sim$4 years of cryostat construction. Thus, we deploy
a temporary softwall cleanroom (CR6) with a higher cleanliness level to serve as the radon-reduced 
work environment for detector installation.

Placed within the confines of CR5, CR6 consists of a sealed structure of thick plastic sheeting on an 
aluminum frame (figure~\ref{fig:CR6}). The dimensions of the enclosure are $\sim$ 3.6 $\times$ 3.6 m$^2$, 
with a ceiling height of $\sim$2.5 m (equal to that of CR5) and an effective volume of $\sim$ 32 m$^3$. 
To lower dust levels, we layed a new stainless steel floor backed 
with plastic sheeting on top of the original tile floor, extending beyond the perimeter of CR6 itself. 
The plexiglass and aluminum door features a floor sweep and an interlocking hand-through port. This door
is aligned with the permanent door between CR5 and CR1. We use the same softwall construction as 
CR6 to set up an enclosed vestibule between the two doors, creating a ``radon lock'' in which
assembly workers wait for air to recirculate before entering the clean area (figure~\ref{fig:CR}).
We find that there is no change in the radon levels in CR6 when personnel remain in the vestibule
for one hour or more before entering the cleanroom, though the increase is acceptably small for a wait
time of about 10 minutes.
A second point of access is a heavy duty zipper in the plastic sheeting, aligned with the large double 
doors leading to the adjacent storage cleanroom (CR4). This can be used for the transport of bulkier 
equipment. All necessary electrical cables, hoses, and nitrogen lines are brought in through sealed 
feedthroughs mounted in metal panels attached to the plastic walls. We take advantage of the 
transparency of the enclosure, mounting the LCD displays of various monitoring devices outside of CR6, 
on the solid walls of CR5. 

The air supply for CR6 is provided by the RAS described previously, in section~\ref{subsec:ras}, through a port
near the ceiling. Two Envirco IsoClean HEPA filter cabinets retrofitted with cleaner Teflon-based 
ULPA filters perform the circulation and filtration of air within CR6. Besides being even more effective
than regular HEPA filters, the ULPA units are radiologically superior---the HEPA filter medium is made
up of randomly arranged fibers spun from borosilicate glass, which is a potential radon contributor. 
We disable CR5's standard air handler, and seal off the HEPA filter outlets in the ceiling with metal 
panels to prevent the entry of radon-containing air. One of the two cabinets is fitted with a heat 
exchanger and connected to an external chiller, helping to keep ambient temperatures lower during work.

We make an effort to render CR6 nearly hermetic to CR5, using cleanroom-standard foam and tape on 
all edges, as well as 3M VHB (Very High Bond) double-sided tape on softwall seams. Combined with 
the disabling of the CR5 air handlers, this sealing ensures that the RAS air flow is sufficient to 
over-pressurize CR6 with respect to its surroundings. We can observe the softwalls ``inflate,''
and that the majority of air current out of CR6 passes through the entry vestibule, allowing it to 
function as designed. 

Beyond the technical capabilities of the RAS and CR6, we can improve cleanliness by adopting certain
work practices. Two particularly important ones during CUORE installation are a daily washing
of the interior surfaces of CR6 as well as tools, and the usage of a nitrogen-flushed 
``radon bag'' (RB). This is a cylindrical plastic barrier hung around the bottom of the cryostat, 
covering those towers that are already mounted. It is continuously flushed with nitrogen, and is 
deployed at all times during which installation is not being performed. While playing a role in protecting 
the detector from recontamination under normal circumstances, the RB is especially important in the 
event of a failure of the RAS.

Once detector installation is complete and the innermost (10 mK) cryostat vessel is closed, we must 
open the hatch in the floor of CR5 to allow the raising of the remaining shields. This is impossible 
to do with CR6 in place. To minimize exposure during this phase, instead of returning
CR5 to its standard configuration, we make use of an ``intermediate'' cleanroom setup. We keep the 
standard HEPA air supply disabled and blocked off, continuing to rely on the RAS and one of the two 
ULPA cabinets. To prevent radon-contaminated air entering from the hall and adjacent cleanrooms, we 
seal the doors and other gaps with cleanroom-rated foam and tape. The vestibule remains in place, 
permitting personnel to enter and exit without flooding the room with external air. 

The conditions inside CR6 are actively monitored.
We use a KNF diaphragm pump to send cleanroom air to the RM described in section~\ref{subsec:rm}. 
Besides the radon level, we also keep track of the airborne particle count, humidity, temperature, 
O$_2$ and CO$_2$ levels. These sensors are equipped with alarms, and at least one shifter is 
watching the readouts at all times.

The softwall modular design of CR6 allows us to deploy or dismount it in a period of less than
one week. This means that we can create a radon-reduced environment around CUORE on short notice
if this becomes necessary in the future.

\section{System performance}
\label{sec:performance}

The achievable sensitivity of the RM measurement depends on the integration time. During the installation
of CUORE, we operated the RM with continuous flushing at a rate of 7 L/min, which gave us a response time 
of $\sim$10 min for $^{218}$Po. This delay primarily stems not from integration, but from the time needed 
to pump the air from CR6 and fill the vessel of the RM, as well as the recirculation time for CR6 itself. 

\begin{figure}[htbp]
  \centering
  \includegraphics[width=0.55\textwidth]{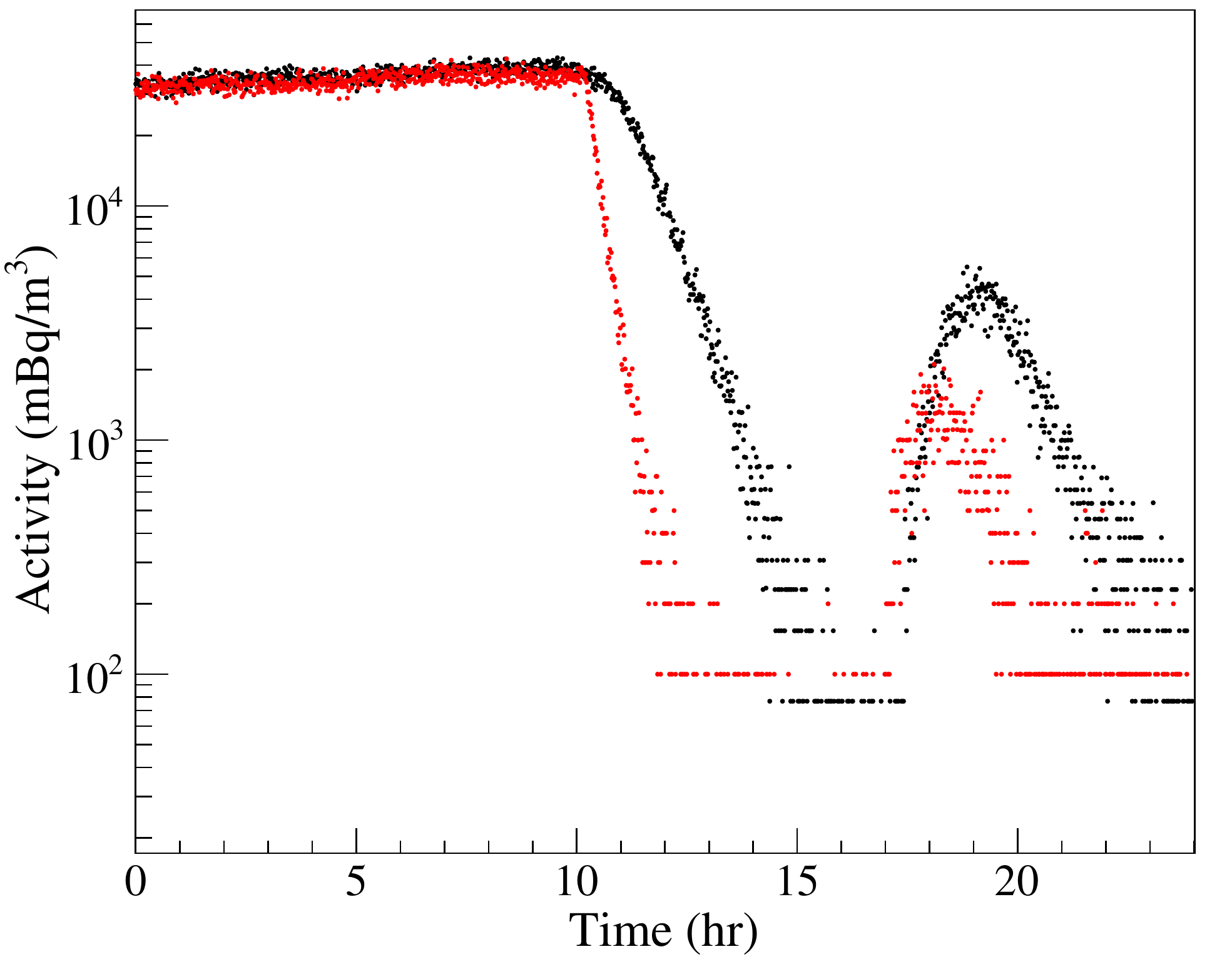}
  \caption{\label{fig:Rn_vs_time}\footnotesize{$^{218}$Po (red) and $^{214}$Po (black) count rate v. time, immediately after 
      the completion of CR6 construction. The first 11 hours represent radon-containing air, 
      while the subsequent measurements are of CR6 air, showing time response 
      to opening and closing CR6.}}
\end{figure}

In Figure~\ref{fig:Rn_vs_time}, we show the activities of $^{218}$Po and $^{214}$Po as a function 
of time, measured immediately after the completion of CR6 construction. The first 11 hours represent 
a readout of dessicated ambient air from the hall, with an average level of $\sim$30 Bq/m$^3$. 
The rapid decrease of $^{ 218}$Po activity down to $\sim$100 mBq/m$^3$ starting around 11:00 
corresponds to switching the measurement to CR6 air. The subsequent increase in count rate around 17:00
occurred during a test where the CR6 door was left open for several minutes. Something we notice
here is the delay of observed $^{214}$Po levels with respect to $^{218}$Po, which occurs due to the 
20--30 minute half-lives of the intermediate $^{214}$Pb and $^{214}$Bi nuclei (figure~\ref{fig:Uchain}).

\begin{figure}[htpb]
  \centering
  \includegraphics[width=0.95\textwidth]{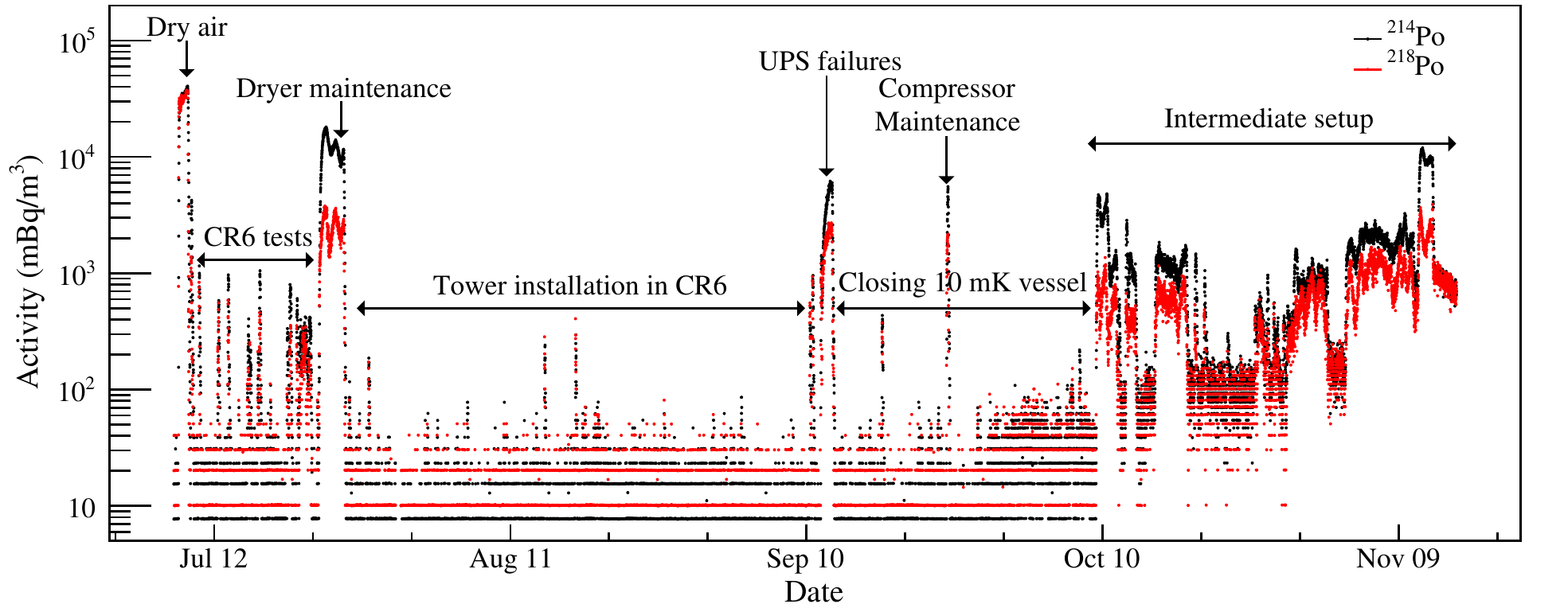}
  \caption{\label{fig:Rn_vs_time_full}\footnotesize{$^{218}$Po (red) and $^{214}$Po (black) count rate v. time over the course of the
      entire $\sim$4 month installation and cryostat closing procedure, with different operation phases labeled. With 
      the ``intermediate'' configuration, 50 mBq/m$^3$ could be stably achieved. The large excursions above that level
      occurring in October and November correspond to the frequent openings of the $\sim$ 2 $\times$ 2 m$^2$ hatch
      in the floor of CR5.}}
\end{figure}

Over the course of the tower installation period, we are able to maintain the radon in CR6
well below the target of 1 Bq/m$^3$ at all times (figure~\ref{fig:Rn_vs_time_full}). When the team is working inside, 
the activity is stable at the $\sim$100 mBq/m$^3$ level. If no activities are being performed, the 
activity decreases to $\sim$10 mBq/m$^3$, not counting the nitrogen-flushing of the RB immediately 
surrounding the detector. When personnel enter the cleanroom under normal working conditions, opening the
door does not cause increases in count rate as large as the one seen in Figure~\ref{fig:Rn_vs_time} thanks to
the use of the vestibule. Reducing the radon level in the empty CR6 from the ambient $\sim$30 Bq/m$^3$ to the 
minimum $\sim$10 mBq/m$^3$ takes approximately 7 hours. We find that the airborne particle count in CR6
remains below 100 particles/ft$^3$, which would correspond to an improvement over the standard
CUORE cleanroom by one class.

In the ``intermediate'' configuration of CR5---with radon-free air from the RAS, a vestibule, and sealed gaps, 
but without the CR6 enclosure---we are able to rach a baseline radon level of $\sim$50 mBq/m$^3$ when not
performing work (figure~\ref{fig:Rn_vs_time_full}), not counting the nitrogen flushing of the 10 mK cryostat vessel. 
While this is five times higher than with CR6 deployed, it is well within our target levels. A quantitative summary of the 
RAS and cleanroom performance is given in Table~\ref{tab:1}.

\begin{table}[htbp]
  \centering
  \caption{\label{tab:1} \footnotesize{CUORE RAS and cleanroom performance summary. Particle count, relative humidity, and ambient 
    temperature were actively monitored only in CR6. CR5 with the floor hatch closed is rated class 1000. 
    Relative humidity in CR6 while unoccupied is stable at 0.5\%, and can reach as high as 25\% during active 
    manual work. Ambient temperature remains at $\sim$19 $^{\circ}$C most of the time.}}
  \smallskip
  \begin{tabular}{|lr|c|c|c|}
    \hline
    &&&Intermediate,&Intermediate,\\
    &&CR6&sealed&open floor hatch\\
    \hline
    Rn: $^{214}$Po and $^{218}$Po&(mBq/m$^3$)& $\sim$10 & 50--100 & 1000--2000\\
    Particle count&(part./ft$^3$)& 0--100 & \textit{rated class 1000} & \textit{not measured}\\
    Relative humidity&(\%)& 0.5--25 & \textit{not measured} & \textit{not measured}\\
    Room temperature&($^{\circ}$C)& 18--23 & \textit{not measured} & \textit{not measured}\\
    \hline
  \end{tabular}
\end{table}

\section{Conclusion}
\label{sec:conclusion}

The CUORE experiment is a state of the art search for the $0\nu\beta\beta$ decay of $^{130}$Te with 
the world's first ton-scale bolometric detector. To achieve our ambitions physics sensitivity targets
with this device, we must ensure its radiopurity. A crucial component of this task is minimizing 
recontamination during the installation of the detector in the cryostat. The combination of the Radon 
Abatement System (RAS) and a dedicated softwall cleanroom enclosure (CR6) allows us to create 
a stable working environment with radon activity in the $\sim 10 - 100$ mBq/m$^3$ range and $<$ 100
airborne particles/ft$^3$ for a period of several months. These performance parameters surpass
our target limits. Using the electrostatic radon monitor (RM), we are able to continuously track
the radon level inside the cleanroom down to values $<$ 10 mBq/m$^3$ with a time delay of 
10 minutes. During CUORE commissioning, we successfully operated the system over the course of 
a four month period, and avoided recontamination of the detector in accordance with our 
background goals. 

\acknowledgments

We would like to thank the staff of the Laboratori Nazionali del Gran Sasso, Lawrence Berkeley National Laboratory, 
and the members of the CUORE Collaboration for valuable assistance in executing the project. 
We are especially grateful to K. Armenta, B. Bohlin, D. Ciccotti, S. Dutta, C. Rusconi, L. Scarcia, J. Schmidt, 
A. Wong, and LNGS staff for their invaluable assistance with servicing CR6, 
as well as carrying out 24/7 monitoring of the sensors and alarms.

This work was supported by the US Department of Energy (DOE) Office of Science under Contract No. DE-AC02-05CH11231, by the DOE Office of Science, Office of Nuclear Physics under Contract No. DE-FG02-08ER41551, and by the National Science Foundation under grant PHY-1314881. The United States Government retains and the publisher, by accepting the article for publication, acknowledges that the United States Government retains a non-exclusive, paid-up, irrevocable, world-wide license to publish or re-produce the published form of this manuscript, or allow others to do so, for United States Government purposes.


\end{document}